\documentclass[conference,letterpaper]{IEEEtran}

\usepackage[utf8]{inputenc} 
\usepackage[T1]{fontenc}
\usepackage{url}
\usepackage{ifthen}
\usepackage{cite}
\usepackage[cmex10]{amsmath} 

\usepackage{textcase}
\usepackage[tablename=Table]{caption}
\usepackage{blindtext}
\usepackage{floatrow}
\usepackage{soul}
\usepackage{changes}
\usepackage{gensymb}
\usepackage{cite}
\usepackage{amsmath,amssymb,amsfonts}
\usepackage{textcomp}
\usepackage{graphicx}
\usepackage{amssymb}
\usepackage{amsfonts}
\usepackage{amsmath}
\usepackage{epsfig}
\usepackage{color}
\usepackage{fancybox}
\usepackage{textcomp}
\usepackage{multirow}
\usepackage{setspa ce}
\usepackage{psfrag}
\usepackage{booktabs}
\usepackage{float}
\usepackage[caption = false]{subfig}
\usepackage{algorithm}
\usepackage{algpseudocode}
\usepackage{mathtools, nccmath, bigints, amsfonts}
\usepackage{mathrsfs}

\newfloatcommand{capbtabbox}{table}[][0.4\textwidth]

\newtheorem{Remark}{Remark}

    \def\Complex{{\rm\rule[.23ex]{.03em}{1.1ex}\kern-.3em{C}}}

    \newcommand{\be}{\begin{equation}} \newcommand{\ee}{\end{equation}}
    \newcommand{\bea}{\begin{eqnarray}} \newcommand{\eea}{\end{eqnarray}}
    \newcommand{\benum}{\begin{enumerate}} \newcommand{\eenum}{\end{enumerate}}

    \newcommand{\qa}{{\bf a}}
        \newcommand{\qb}{{\bf b}}
        \newcommand{\qc}{{\bf c}}

        \newcommand{\qn}{{\bf n}}

        \newcommand{\qq}{{\bf q}}

        \newcommand{\qv}{{\bf v}}
        \newcommand{\qw}{{\bf w}}
        \newcommand{\qx}{{\bf x}}
        \newcommand{\qy}{{\bf y}}
        \newcommand{\qz}{{\bf z}}
    
        \newcommand{\qA}{{\bf A}}
        \newcommand{\qB}{{\bf B}}
        \newcommand{\qC}{{\bf C}}
        \newcommand{\qD}{{\bf D}}

        \newcommand{\qH}{{\bf H}}
        \newcommand{\qI}{{\bf I}}

        \newcommand{\qV}{{\bf V}}

        \newcommand{\qzero}{{\bf 0}}

        \newcommand{\qSigma}{{\boldsymbol \Sigma}}

        \newcommand{\qOmega}{{\boldsymbol \Omega}}

        \newcommand{\qlambda}{{\boldsymbol \lambda}}
        \newcommand{\qgamma}{{\boldsymbol \gamma}}

        \newcommand{\qmu}{{\boldsymbol \mu}}

        \newcommand{\calN}{{\mathcal N}}

        \newcommand{\Ex}{{\sf E}}

\def\BibTeX{{\rm B\kern-.05em{\sc i\kern-.025em b}\kern-.08em
    T\kern-.1667em\lower.7ex\hbox{E}\kern-.125emX}}
\begin{document}

\title{Improving Cell-Free Massive MIMO Detection Performance via Expectation Propagation } 


 \author{%
   \IEEEauthorblockN{Alva Kosasih\IEEEauthorrefmark{1},
                     Vera Miloslavskaya\IEEEauthorrefmark{1},
                     Wibowo Hardjawana\IEEEauthorrefmark{1},
                     Victor Andrean\IEEEauthorrefmark{2},
                     and Branka Vucetic\IEEEauthorrefmark{1}}
   \IEEEauthorblockA{\IEEEauthorrefmark{1}%
                     Centre of Excellence in Telecommunications, University of Sydney, Sydney, Australia.  \\
                    \{alva.kosasih,vera.miloslavskaya,wibowo.hardjawana,branka.vucetic\}@sydney.edu.au  }
   \IEEEauthorblockA{\IEEEauthorrefmark{2}%
                     Mobilizing Information Technology Lab., National Taiwan University of Science and Technology, Taipei, Taiwan. }
 }

\maketitle

\begin{abstract}
   
Cell-free (CF) massive multiple-input multiple-output (M-MIMO) technology plays a prominent role in the beyond fifth-generation (5G) networks. However, designing a high performance CF M-MIMO detector is a challenging task due to the presence of pilot contamination which appears when the number of pilot sequences is smaller than the number of users. This work proposes a CF M-MIMO detector referred to as CF expectation propagation (CF-EP) that incorporates the pilot contamination when calculating the posterior belief. The simulation results show that the proposed detector achieves  significant improvements in terms of the  bit-error rate and sum spectral efficiency performances as compared to the ones of the state-of-the-art CF detectors. 
\end{abstract}



\section{Introduction}

Cell-free (CF) massive multiple-input multiple-output (M-MIMO) technology has been proposed to deliver ubiquitous coverage and high service quality for  beyond fifth-generation (5G) networks \cite{2020JZhang_JSAC_ProsMulAntennas}. The access points (APs), known as the base station antennas in standard MIMO literature, are geographically distributed. The APs are managed by a central processing unit (CPU).
 Each user equipment (UE) is served simultaneously by all APs and therefore experiences no cell boundaries. This CF M-MIMO system provides additional diversity, increases the coverage probability, and mitigates the inter-cell interference, leading to a higher spectral efficiency (SE) and  energy efficiency \cite{2017HQNgo_TWC_CFvsSmall,2020EBjornson_TWC_MakingCFcompetitive,2019YFeng_ICC_IterDetDMIMO}.  One of the critical issues is to design a detector for CF M-MIMO system which can overcome various impairments, in particular the pilot contamination,  i.e. interference caused by pilot-sharing UEs \cite{2020EBjornson_TWC_MakingCFcompetitive}.   

 The performance of the classical maximum ratio combining (MRC) detector in a CF M-MIMO system, referred to as CF-MRC detector,  has been investigated in \cite{2017HQNgo_TWC_CFvsSmall}. The CF-MRC detector estimates the transmitted symbols by multiplying corresponding received signals with the complex conjugate of the channel matrix. Although the CF-MRC detector can achieve an extremely low computational complexity, it fails to overcome the  pilot contamination impact, which leads to a severe performance degradation   \cite{2020EBjornson_TWC_MakingCFcompetitive}. 
 To deal with the pilot contamination in CF M-MIMO systems,  a  modified minimum mean square error (MMSE) detector, referred to as CF-MMSE, has been proposed in \cite{2016ENayebi_Asilomar_CFMMSE}. It is shown to  significantly outperform the CF-MRC detector by incorporating the statistical channel estimation error which contains the pilot contamination term when detecting the symbols. Specifically, the variance of the channel estimation error is incorporated into the regularization factor in case of the CF-MMSE detector.  To further improve the performance of the CF-MMSE detectors, interference cancellation techniques have been combined with the CF-MMSE detectors as proposed in \cite{2019YFeng_ICC_IterDetDMIMO,2020EBjornson_TWC_MakingCFcompetitive,2020RMosayebi_TWC_LinearInterference}. 
It is widely known that the MMSE with successive interference cancellation (SIC) detectors can  achieve a high reliability performance at the cost of a higher computational complexity compared to the standard MMSE detector  \cite{2008TLiu_TWC_MMSESIC}.  
The CF-MMSE-SIC detector, combining the CF-MMSE scheme with the SIC scheme has been proposed in \cite{2020EBjornson_TWC_MakingCFcompetitive}. Unfortunately, the CF-MMSE-SIC detector provides only a minor performance gain compared to the CF-MMSE detector. 

 
Recently, Bayesian machine learning techniques, called expectation propagation (EP) \cite{Jespedes-TCOM14} and approximate message passing (AMP) \cite{2011Bayati_AMP}, have been employed for detection to reduce the reliability performance gap between ML and suboptimal detectors in the collocated M-MIMO systems. In \cite{2011Bayati_AMP}, the AMP algorithm has been shown to achieve a near optimal performance for independent and identically distributed (i.i.d.) Rayleigh fading channels. However, it performs poorly for non-i.i.d Rayleigh fading channels \cite{Rangan2017VAMP,Takeuchi2017EP}.
The EP algorithm is an iterative algorithm used to infer estimates of the transmitted symbols by approximating their posterior probability density function (pdf) with factorizable Gaussian posterior belief \cite{Jespedes-TCOM14}. For each posterior belief, EP retains only expectations such as mean and variance. The factorizable belief not only greatly reduces the complexity of the posterior pdf inference, but also enables the EP to achieve a near optimal performance \cite{2020AKosasih_WCNC_LinearBayesian}.

{We propose a novel iterative CF M-MIMO detector based on the EP algorithm, referred to as the CF-EP detector.  The main contributions are summarized as follows: 
\begin{itemize}
\item To the best of our knowledge, this is the first Bayesian detector for  CF M-MIMO systems which can achieve a high detection reliability in the presence of the pilot contamination. 
\item The Gaussian posterior belief calculation of the original EP detector \cite{Jespedes-TCOM14} is modified for systems with the pilot contamination, which is the main performance degradation factor in CF M-MIMO systems. 
\end{itemize}
The simulation results demonstrate a significant BER and SE performance improvements compared to the state-of-the-art  \cite{2017HQNgo_TWC_CFvsSmall,2016ENayebi_Asilomar_CFMMSE,2019YFeng_ICC_IterDetDMIMO,2020EBjornson_TWC_MakingCFcompetitive} with a comparable computational complexity.}

{\bf Notations}: $\mathbf{I}$ denotes an identity matrix.  For any matrix $\mathbf{A}$, the notations $\mathbf{A}^{T}$, $\mathbf{A}^{H}$, ${\rm tr}(\mathbf{A})$, and $\mathbf{A}^{\dagger}$ stand for transpose, conjugate transpose, trace, and pseudo-inverse of $\mathbf{A}$, respectively.  $\|\qq\|$ denotes the Frobenius norm of vector $\qq$.   $q^*$ denotes the complex conjugate of a complex number $q$. 
 ${\Ex}[\qx]$ is the mean of random vector $\qx$, and ${\mathrm{Var} }[\qx] = {\Ex}\big[\left(\qx-{\Ex}[\qx]\right)^2\big]$ is its variance.   $\calN_{\mathbb{C}}(x_k: c_k,v_k)$ represents a complex single variate Gaussian distribution  for a  random variable $x_k$ with mean $c_k$ and variance $v_k$. Let $\qx = [x_1, \cdots, x_K]^T$ and $\qc = [c_1, \cdots, c_K]^T$.

\section{System Model}

We consider an  uplink CF M-MIMO system which consists of $L$ geographically distributed  single-antenna APs, serving $K$ single antenna UEs, where $K << L$. All APs send their pilot and data signals  to the CPU which then estimates the channel state information (CSI) and transmitted symbols. We use a block fading model, where the channel between the $k$-th UE and the $l$-th AP is expressed as
\begin{equation}\label{eq_SM:large_scale}
h_{l,k}  =\sqrt{\beta_{l,k}}g_{l,k}.
\end{equation}
Here,  $g_{l,k} \sim \mathcal{N}_\mathbb{C} \left( 0,1\right)$ is the complex small scale fading coefficient and $\beta_{l,k} $ is the complex large-scale fading coefficient that describes geometric pathloss and shadowing.  We assume that  $\beta_{l,k} $ is known to the CPU. The channel is considered to be constant in time-frequency blocks of $\tau_c$ channel uses. Note that precise value of $h_{l,k}$ is unavailable to the receiver. Therefore, the channel is estimated by using the minimum mean-square error criterion, i.e. minimization of $\Ex [ \| h_{l,k} - \hat{h}_{l,k} \|^2] $,  where $\hat{h}_{l,k}$ denotes the MMSE channel estimate of UE $k$ from AP $l$. This results in an optimal estimation quality \cite{2017Bjornson_BOOK_MMIMO}.

 \begin{figure}
\centering
{\includegraphics[scale=0.28]{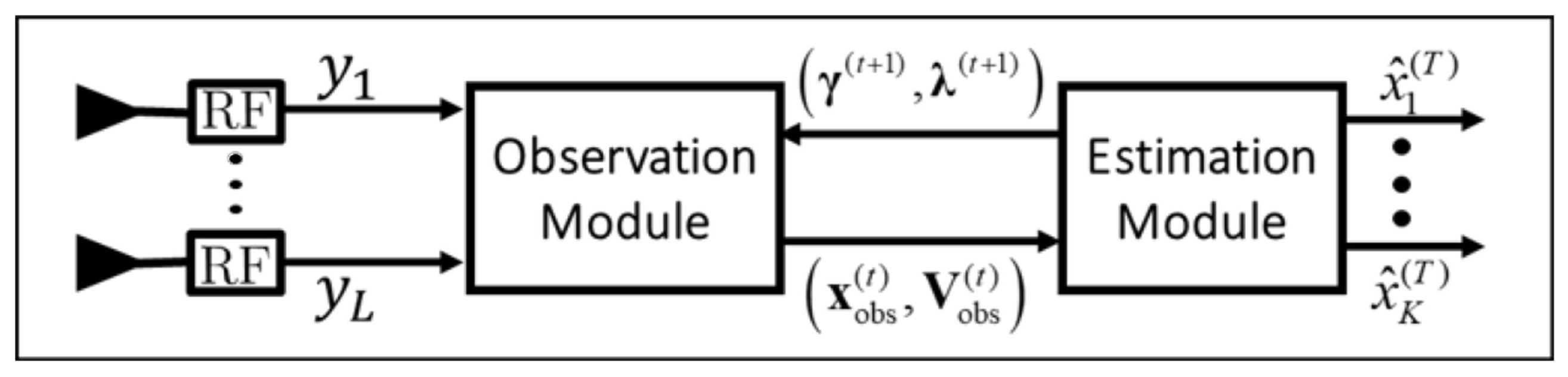}}\hfill
\caption{ Cell-free expectation propagation (CF-EP) detector}
\label{F1}
\end{figure} 

 \subsection{Uplink Pilot Transmission and MMSE Channel Estimation}
 
 The channels between UEs and APs are estimated by using $\tau$ mutually orthogonal pilot signals $\boldsymbol{\phi}_1, \dots, \boldsymbol{\phi}_{\tau}$ with $ \boldsymbol{\phi}_{i}\in\mathbb{C}^{\tau\times 1}, \|\boldsymbol{\phi}_{i}\|^2=\tau, $ and $ i = 1,\dots,\tau $. The pilots are randomly assigned to the UEs. In practice, $\tau < K $ which results in a pilot contamination \cite{2017HQNgo_TWC_CFvsSmall}. 
 Let $t_k$ be the index of the pilot signal assigned to the $k$-th UE, $1\leq k\leq K$. Obviously, $t_k\in\{ 1,\dots,\tau\}$. 
 We group UEs into $\tau $ sets $\mathcal{S}_1, \dots, \mathcal{S}_\tau$ such that all UEs in the set use the same pilot, i.e. $\mathcal{S}_i\triangleq\{ k\in\{ 1,\dots,K\} | t_k=i \}$ for $1\leq i\leq\tau$. At AP $l$, the received pilot  $\qz_l \in \mathbb{C}^{\tau \times 1}$ is given as follows,
\begin{equation}\label{eq_SM:pilot_trans}
\qz_l = \sqrt{p} \sum_{i=1}^K h_{l,i} \boldsymbol{\phi}_{t_i} + \qv_l,
\end{equation}
where $p$ is the uplink transmit power of a UE\footnote{All UEs are assumed to transmit with equal power.} and  $\qv_l  \in \mathbb{C}^{\tau \times 1} \sim \mathcal{N}_\mathbb{C}  (\bold{0},\sigma^2 \qI)$ is an additive Gaussian noise with zero mean and variance $\sigma^2$ equal to the noise power.

The MMSE channel estimate of UE $k$ from AP $l$ is given in \cite{2020EBjornson_TWC_MakingCFcompetitive} as
\begin{equation}\label{eq_SM:MMSE_Ch_estim}
\hat{h}_{l,k} = \frac{\sqrt{p \tau }\beta_{l,k}}{\sigma^2+p \tau \sum_{i \in \mathcal{S}_{{{t_k}}}} \beta_{l,i}} \times \frac{ {\boldsymbol{\phi}_{t_k}^H}}{\sqrt{\tau}}  \qz_l  
\end{equation}
By substituting $\qz_l $ in  \eqref{eq_SM:pilot_trans} into \eqref{eq_SM:MMSE_Ch_estim}, we obtain
\begin{equation}\label{eq_SM:expandMMSE_Ch_estim}
\hat{h}_{l,k}= \frac{\sqrt{p \tau }\beta_{l,k}}{\sigma^2+p \tau \sum_{i \in \mathcal{S}_{{{t_k}}}} \beta_{l,i}} \times \left( \sqrt{p \tau }  \sum_{i \in \mathcal{S}_{t_k}} h_{l,i} + \tilde{\qv}_l \right),
\end{equation}
where each element in $\tilde{\qv}_l$ follows a Gaussian distribution with zero mean and variance $\sigma^2$.
From \eqref{eq_SM:expandMMSE_Ch_estim} and \eqref{eq_SM:large_scale}, we can easily verify that  $ \Ex \left[ \hat{h}_{l,k}\right] =0$ and $ {\alpha_{l,k} \triangleq \Ex } \left[ \hat{h}_{l,k} (\hat{h}_{l,k})^* \right]  = \frac{p \tau\beta_{l,k}^2}{\sigma^2+p \tau \sum_{i \in \mathcal{S}_{{t_k}}} \beta_{l,i}}   $, $\hat{h}_{l,k}  \sim \mathcal{N}_{\mathbb C} \left( 0, \alpha_{l,k} \right)$.
Therefore, the channel estimation error of UE $k$ from AP $l$ 
 \begin{equation}\label{eq_SM:Ch_estim_err}
 \epsilon_{l,k} = h_{l,k} - \hat{h}_{l,k}
\end{equation}  
 follows a Gaussian distribution with mean zero and variance  $C_{l,k} \triangleq  \Ex [ {\epsilon}_{l,k} ({\epsilon}_{l,k})^*]  = \beta_{l,k} - \alpha_{l,k} $. 
 The channel estimation error from the $k$-th UE to all APs is defined in a vector form as $\boldsymbol{\epsilon}_k\triangleq [\epsilon_{1,k},\dots, \epsilon_{L,k}]^T$, where $\boldsymbol{\epsilon}_k \sim \mathcal{N}_{\mathbb C} \left( 0, \qC_{k} \right)$. The covariance matrix $\qC_k$ is an $L \times L$ diagonal matrix whose diagonal elements are $C_{1,k}, \dots, C_{L,k}$. 
 The channel estimation error matrix for $K$ UEs, $ \boldsymbol{\mathcal{E}} \triangleq [\boldsymbol{\epsilon}_{1}, \cdots, \boldsymbol{\epsilon}_{K}] $ can be written as
\begin{equation}\label{eq_SM:Ch_estim_err_matrix}
\boldsymbol{\mathcal{E}}= \qH - \hat{\qH}.
\end{equation}  
 Since $ {\epsilon}_{l,k}  \sim \mathcal{N}_{\mathbb C} \left( 0, C_{l,k} \right)$, the channel estimation error matrix  $ \boldsymbol{\mathcal{E}}  $ is normally distributed with zero mean and covariance matrix    $\qD \triangleq \Ex\left[ \boldsymbol{\mathcal{E}} (\boldsymbol{\mathcal{E}})^H  \right] = \sum_{k=1}^K \qC_k$, i.e. $\boldsymbol{\mathcal{E}}\sim\mathcal{N}(\mathbf{0}, \mathbf{D})$.

\subsection{Uplink Data Transmission}

All UEs map their information bit streams to symbols that belong to constellation of $M$-QAM, $\qOmega$, where $\|\qOmega\|^2=M$. The $k$-th user symbol is denoted as $x_k$. The average symbol energy is $E_x \triangleq \Ex[ |x_k|^2]=1$. The channel matrix between UEs and APs and its estimate are denoted as  $\qH  \in \mathbb{C}^{L\times K} $ and $\hat{\qH}  \in \mathbb{C}^{L\times K} $, respectively. The $(l,k)$-th elements of $\qH$ and $\hat{\qH}$ are defined by \eqref{eq_SM:large_scale} and \eqref{eq_SM:MMSE_Ch_estim}, respectively. 
The received signal of all APs, $\bold{y}=[y_1 \ldots y_L]^{T}$, available at the CPU, is given as
\begin{equation} \label{eq_SM:RecSigRaw}
\qy = \qH \qx + \qn, 
\end{equation}
where  $\qn  \sim \mathcal{N}_\mathbb{C} (\bold{0}, \sigma^2 \qI)$ is a Gaussian noise and $\qx= [x_1, \cdots, x_K]^T$ is uniformly distributed on $\qOmega^K$.
By substituting $\hat{\qH}$ in \eqref{eq_SM:Ch_estim_err_matrix} into \eqref{eq_SM:RecSigRaw}, we obtain
\begin{equation}\label{eq_SM:RecSig}
\qy =\hat{\qH} \qx + \underbrace{ \boldsymbol{\mathcal{E}} \qx   + \qn}_{\qw}, 
\end{equation}
 at the CPU. To detect the symbols we can use the MMSE channel estimate $\hat{\qH}$,  which introduces the channel estimation error $ \boldsymbol{\mathcal{E}} $ as explained in the Section IIA.
Considering that $\Ex \left[   \boldsymbol{\mathcal{E}}  \qx   + \qn  \right] = \bold{0}$ and $\mathrm{Var} \left[  \boldsymbol{\mathcal{E}}  \qx   + \qn  \right] = \qD+  \sigma^2 \qI $, we define a new noise vector $\qw \sim \mathcal{N}_\mathbb{C} (\bold{0},  \qD+  \sigma^2 \qI)$.


\section{Cell-Free Expectation Propagation Detector}

In this section, we propose a novel CF M-MIMO detector based on the EP concept \cite{Jespedes-TCOM14}. Our detector takes the effect of the pilot contamination into consideration. We refer it as the cell-free expectation propagation (CF-EP) detector with the architecture  shown in Fig. \ref{F1}. The CF-EP consists of two modules: the observation module calculating the posterior beliefs based on the received signal and the estimation module yielding soft transmitted symbol estimates. The CF-EP iteratively exchanges the outputs of both modules until the convergence criterion is satisfied.

\subsection{The Observation Module}

Given the received signal $\qy$  in \eqref{eq_SM:RecSig}, the posterior distribution of the transmitted symbols is characterized by
\begin{flalign} \label{eq_EP:Posterior_ori}
    & p(\qx|\qy) = \frac{p(\qy|\qx) }{ p(\qy)} \times p(\qx) \propto \underbrace{\mathcal{N}_\mathbb{C} \left( \qy: \hat{\qH} \qx , \qD+  \sigma^2 \qI \right) }_{p(\qy|\qx)}    \underbrace{ \prod_{k=1}^{K} p(x_{k}) }_{p(\qx)},
\end{flalign}
where  $p(x_k) = \frac{1}{M} \sum_{x \in \qOmega } \delta(x_k-x)$ is a priori pdf of $x_k$, $\delta$ is the Dirac delta function, and $p(\qy)$ is omitted as it does not depend on the distribution of $\qx$.  
A direct calculation of \eqref{eq_EP:Posterior_ori} results in a prohibitively intensive computation {as we need to consider all possible combinations of the symbols}. 
Therefore, EP is used to iteratively construct a Gaussian approximation to the true posterior distribution of the transmitted symbol vector. 

Concretely, the EP constructs a Gaussian posterior belief $ p^{(t)}(\qx|\qy)\approx p(\qx|\qy) $, where $t$ is the iteration number.
This involves replacing $p(\qx)$ in \eqref{eq_EP:Posterior_ori}  with a distribution from the exponential
family, $\chi^{(t)}(\qx)\propto \mathcal{N}_\mathbb{C} \left(\qx: (\qlambda^{(t)})^{-1} \qgamma^{(t)}, (\qlambda^{(t)})^{-1}  \right)$ \cite{2020JZhang_Arxiv_MetaLearning}. 
Here, $\qlambda^{(t)} $ is a $K \times K$ diagonal matrix with diagonal elements $\lambda_{k}^{(t)}>0 $  and $ \qgamma^{(t)} = [\gamma_{1}^{(t)}, \dots, \gamma_{K}^{(t)}]^T$. Both  $\lambda_{k}^{(t)}$ and  $\gamma_k^{(t)}$  are complex numbers with  $\lambda_{k}^{(0)}=1$ and $\gamma_k^{(0)}=0$. { To calculate the Gaussian posterior belief, we first treat $\qx$ as a random vector and approximate $p(\qy|\qx) $ as $ \mathcal{N}_\mathbb{C}  \left( \qx: {\hat{\qH}}^\dagger \qy , \left( \hat{\qH}^H (\qD+  \sigma^2 \qI)^{-1} \hat{\qH} \right)^{-1} \right) $. Note that the matrix $\qD$ corresponds to the channel estimation error variance which is caused by the pilot contamination, defined in Section IIA. The Gaussian posterior belief is then expressed as }
\begin{flalign}\label{eq_EP:Post_approx}
p^{(t)}(\qx|\qy) \propto & p(\qy | \qx) \cdot \chi^{(t)}(\qx)\notag \\
 \propto & \mathcal{N}_\mathbb{C}  \left( \qx: {\hat{\qH}}^\dagger \qy, \left( \hat{\qH}^H (\qD+  \sigma^2 \qI)^{-1} \hat{\qH} \right)^{-1} \right)  \notag \\
&\cdot \mathcal{N}_\mathbb{C}  \left(\qx: (\qlambda^{(t)})^{-1} \qgamma^{(t)}, (\qlambda^{(t)})^{-1}  \right)\notag \\
\propto &\mathcal{N}_\mathbb{C}  \left( \qx:\qmu^{(t)}, \qSigma^{(t)} \right).
\end{flalign}
We can compute the product of two Gaussians in \eqref{eq_EP:Post_approx} by using the Gaussian product property\footnote{The product of two Gaussians results in another Gaussian, $\mathcal{N}_{\mathbb{C}}(\qx:\qa,\qA) \cdot \mathcal{N}_{\mathbb{C}}(\qx:\qb,\qB)  \propto \mathcal{N}_{\mathbb{C}} (\qx:(\qA^{-1}+\qB^{-1})^{-1}(\qA^{-1} \qa + \qB^{-1} \qb),(\qA^{-1}+\qB^{-1})^{-1}$.}, given in Appendix A.1 of \cite{Rasmussen-BOOK}. By using this property, we obtain the variance and mean of $p^{(t)}(\qx|\qy) $ as
\begin{subequations} \label{eA1_a0102}
            \begin{align}
&\qSigma^{(t)} =  {\left(\hat{\qH}^H (\qD+  \sigma^2 \qI)^{-1} \hat{\qH}+ \qlambda^{(t)} \right)}^{-1} \label{eA1_a01}\\
& \qmu^{(t)} =\qSigma^{(t)} {\left(\hat{\qH}^H  (\qD+  \sigma^2 \qI)^{-1} \qy + \qgamma^{(t)}\right)}. \label{eA1_a02}
            \end{align}
        \end{subequations}

\begin{Remark}[CF-EP Gaussian posterior belief]
The Gaussian posterior belief $p^{(t)}(\qx|\qy)$  in the CF-EP detector  is different from that in the existing EP detector, i.e. equation (27) in \cite{Jespedes-TCOM14}, as the channel estimation error containing the pilot contamination term is incorporated  in the proposed CF-EP detector, specifically in \eqref{eq_EP:Post_approx}. 
\end{Remark}

 {We then compute $p^{(t)}(\qy|\qx)$ based on $p^{(t)}(\qx|\qy) $,
 \begin{flalign}\label{eA1_a0304raw}
&p^{(t)}(\qy|\qx)  \triangleq \frac{p^{(t)}(\qx|\qy)}{\chi^{(t)}(\qx)}   \notag\\
&\propto\frac{\mathcal{N}_\mathbb{C}  \left( \qx:\qmu^{(t)}, \qSigma^{(t)} \right)}  {\mathcal{N}_\mathbb{C} \left(\qx:(\qlambda^{(t)})^{-1} \qgamma^{(t)}, (\qlambda^{(t)})^{-1} \right)}  \propto \mathcal{N}_\mathbb{C}  \left( \qx: \qx^{(t)}_{{\rm obs}}, \qV^{(t)}_{{\rm obs}} \right),
 \end{flalign}
 where $ \qx^{(t)}_{\rm obs} = [x^{(t)}_{{\rm obs},1}, \dots,x^{(t)}_{{\rm obs},K}] $ and   $\qV^{(t)}_{\rm obs}$ is a $K \times K$ diagonal matrix with  $v^{(t)}_{ {\rm obs},k}$ as the $k$-th diagonal element, which can be expressed as
 \begin{subequations} \label{eA1_a0304}
            \begin{align}
&v_{{\rm obs},k}^{(t)} =  \frac{\Sigma_{k}^{(t)} }{1- \Sigma_{k}^{(t)}  \lambda_{k}^{(t)}}  \label{eA1_a03}\\
&x_{{\rm obs},k}^{(t)}  = v_{{\rm obs},k}^{(t)}  {\left(\frac{\mu_{k}^{(t)}}{\Sigma_{k}^{(t)}}-\gamma_{k}^{(t)}\right)}.  \label{eA1_a04} 
            \end{align}
        \end{subequations}
Here, $\mu_k^{(t)}$ is the $k$-th element of vector $\qmu^{(t)}$ and $\Sigma_k^{(t)}$ is the $k$-th diagonal element of matrix $\qSigma^{(t)}$.}
The pair $\left(\qx_{{\rm obs}}^{(t)},\qV_{{\rm obs}}^{(t)}\right)$ from \eqref{eA1_a0102}  is then forwarded to the estimation module.

 \subsection{The Estimation Module}    

Based on $p^{(t)}(\qy|\qx)$ provided by the observation module, the estimation module computes a new posterior belief
\begin{equation}\label{eqEP:New_post}
\hat{p}^{(t)}(\qx|\qy) \propto p^{(t)}(\qy|\qx) p(\qx). 
\end{equation}
 This is different from the Gaussian posterior belief $p^{(t)}(\qx|\qy)$ in \eqref{eq_EP:Post_approx}, obtained based on $\chi^{(t)}(\qx)$ from exponential families. In fact, the EP algorithm is used to ensure a similarity between the mean-variance pairs of $p^{(t)}(\qx|\qy)$ and $\hat{p}^{(t)}(\qx|\qy)$. This similarity is specified in \cite{Minka-01} by the moment matching (MM) condition, minimizing the Kullback-Leibler divergence between posterior beliefs. The MM  condition is expressed as follows
\begin{align}\label{eqEP:MM}
\mathbb{E}_{p^{(t)}(\qx|\qy)} [\qx] \sim \mathbb{E}_{\hat{p}^{(t)}(\qx|\qy)} [\qx], 
\end{align}
where $\mathbb{E}_{\hat{p}^{(t)}(\qx|\qy)} [\qx]$ denotes the first two moments, i.e. the mean and variance of $\qx$ with respect to $\hat{p}^{(t)}(\qx|\qy)$. Furthermore, the mean and variance of $\hat{p}^{(t)}(\qx|\qy) [\qx]$ are given in \cite{Jespedes-TCOM14} as
\begin{subequations}\label{eA1_b0102}
            \begin{equation}\label{eA1_b02}
\qV^{(t)} =  \Ex  \left[ \left| \qx  -\hat{\qx}^{(t)}\right|^{2} \right]
            \end{equation}
            \begin{equation}\label{eA1_b01}
\hat{\qx}^{(t)}= c^{(t)}  \sum_{\qx \in \qOmega^K}  \qx  p^{(t)}(\qy|\qx).
            \end{equation}
\end {subequations}
 Here, $c^{(t)}$ is the normalization constant to ensure that $\sum_{\qx \in \qOmega^K} \hat{p}^{(t)}(\qx|\qy)= 1$.
 The iteration of the CF-EP detector will be terminated when
\begin{equation}\label{conv_crit}
 \| \mathbb{E}_{p^{(t)}(\qx|\qy)} [\qx] - \mathbb{E}_{\hat{p}^{(t)}(\qx|\qy)} [\qx]\| \leq 10^{-4}, 
\end{equation} 
i.e. when \eqref{eqEP:MM} is satisfied, or once the algorithm once the maximum number of iterations  $T_{\rm max}$ is reached. Hard estimates of the transmitted symbols are then made  from $\hat{\qx}^{(T)}$, by comparing its Euclidean distance from the transmitted $M$-QAM symbol sets $\Omega$ where $T$ is the last iteration number. 

 \begin{figure}
\centering
{\includegraphics[scale=0.23]{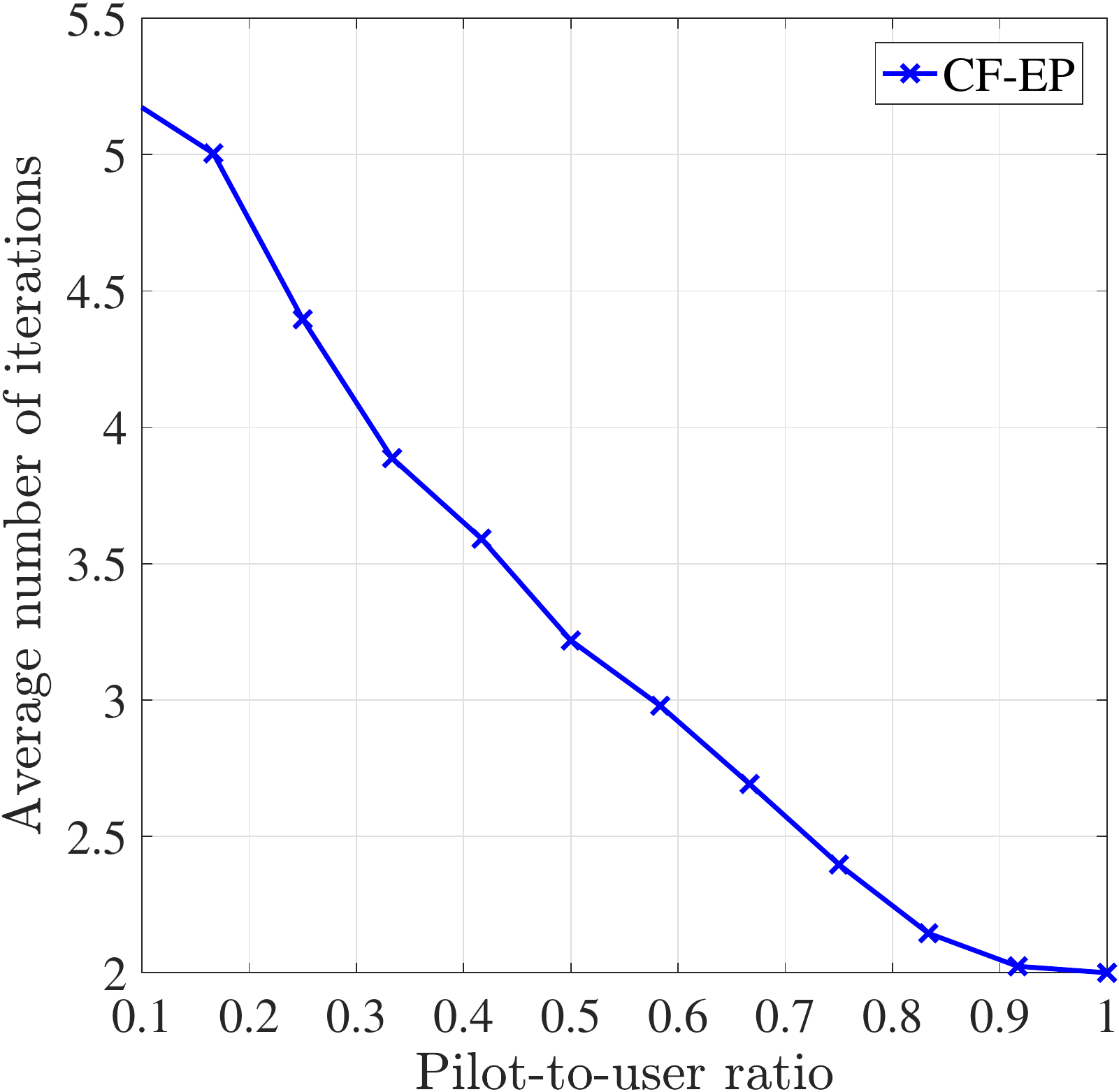}}\hfill
\caption{The average number of iterations  of the CF-EP detector}
\label{F2}
\end{figure}

Until \eqref{eqEP:MM} is satisfied, the estimation module will re-evaluate the Gaussian posterior belief, ${p}^{(t)}(\qx|\qy)$,  by first calculating $\chi^{(t+1)}(\qx)$ based on the $\hat{p}^{(t)}(\qx|\qy)$, expressed as
 \begin{flalign} \label{eq:EP_inference_recon}
\chi^{(t+1)}(\qx) & \triangleq \frac{\hat{p}^{(t)}(\qx|\qy)}{p^{(t)}(\qy|\qx)} \notag \\
 &\propto  \frac{\mathcal{N}_\mathbb{C}  \left( \qx:  \hat{\qx}^{(t)}, \qV^{(t)} \right)}{\mathcal{N}_\mathbb{C} \left( \qx: \qx^{(t)}_{{\rm obs}},\qV^{(t)}_{{\rm obs}}\right) } \notag \\
 &=  \mathcal{N}_\mathbb{C}  \left(  \qx: (\qlambda^{(t+1)})^{-1} \qgamma^{(t+1)},  (\qlambda^{(t+1)})^{-1}   \right),
\end{flalign} 
where
\begin{subequations} \label{eA1_b0304}
            \begin{align}
&\qlambda^{(t+1)} = (\qV^{(t)})^{-1} -  (\qV^{(t)}_{{\rm obs}})^{-1}  \label{eA1_b03}\\
&\qgamma^{(t+1)} =  (\qV^{(t)})^{-1} \hat{\qx}^{(t)} - (\qV^{(t)}_{{\rm obs}})^{-1}   \qx^{(t)}_{{\rm obs}}. \label{eA1_b04}
            \end{align}
\end{subequations}
Note that $\qlambda$ and $\qgamma$ are referred to as the updating parameters as they are used to update the Gaussian posterior beliefs. 
We smoothen the update of $(\qlambda,\qgamma)$ by using a convex combination with the former values,
\begin{subequations} \label{eq:damping}
\begin{align}
   \qlambda^{(t+1)} &= (1-\eta)\qlambda^{(t+1)}+\eta \qlambda^{(t)} \label{damping_lambda} \\
   \qgamma^{(t+1)} &= (1-\eta)\qgamma^{(t+1)}+\eta\qgamma^{(t)}\label{damping_gamma},
\end{align}
\end{subequations}
where $\eta\in[0, 1]$ is a weighting coefficient.

The estimation module sends the updating parameters $(\qgamma^{(t+1)},\qlambda^{(t+1)})$ to the observation module, as illustrated in Fig. \ref{F1}.
The complete pseudo-code is shown in Alg. \ref{A1}. One can also compute the SE of the CF-EP detector by using  signal-to-interference-noise ratio,  given in\cite{1998EBiglieri_InfTheo_SE_SINR} as
\begin{equation}
{\rm SE}^{\rm CF-EP}_k = \left(1 - \frac{\tau_p}{\tau_c} \right) \Ex \left[  {\sf log}_2  \left(  1 +  {{\rm SINR}_k^{\rm CF-EP}}  \right) \right],
\end{equation}
where  ${{\rm SINR}_k^{\rm CF-EP}} = \frac{K}{{\sf tr}\left(\qV_{\rm obs}^{(T)} \right)}$.

\begin{algorithm}
\caption{CF-EP detector }
\label{A1}
\begin{algorithmic}[1]
\State {\textbf{Input: } $\qgamma^{(0)} = \qzero ,  \qlambda^{(0)} = \textbf{I} , \eta = 0.7, t=0, T_{\rm max}=10$}
\State {\textbf{Output: }  Hard symbol estimates from $\hat{\qx}^{(t)}$}
	\For {$t=1,\dots,T_{\rm max}$}
	
	    \Statex \textbf{\quad\, The Observation Module:}
		\State{Compute  $\qSigma^{(t)}$ and $\qmu^{(t)}$  in \eqref{eA1_a0102}}
		\State{Compute  $v_{{\rm obs},k}^{(t)}$  and $x_{{\rm obs},k}^{(t)}$ in \eqref{eA1_a0304}, $k=1,\dots,K$}
		
		\Statex \textbf{\quad\, The Estimation Module:}
		\State{Compute $\qV^{(t)}$ and $\hat{\qx}^{(t)}$ in \eqref{eA1_b0102}}
		\State{Compute  $\qlambda^{(t+1)}$ in \eqref{eA1_b03} and smoothen it using \eqref{damping_lambda}}
		\State{Compute  $\qgamma^{(t+1)}$ in \eqref{eA1_b04} and smoothen it using \eqref{damping_gamma}}
		\If {\eqref{conv_crit} is satisfied}
			\State {break}
		\EndIf
	\EndFor
\State $T: =t$ 
\State Calculate hard symbol estimates from $\hat{\qx}^{(T)}$
\end{algorithmic}
\end{algorithm}

\section{Numerical Results}\label{Sim_configs}

{We follow the simulation setup used in \cite{2020EBjornson_TWC_MakingCFcompetitive}}. Specifically, we consider $L = 100$ single-antenna APs, randomly deployed in a  $1 \times 1$ km area, serving $K = 60$ randomly located UEs, in an urban environment complying with 3GPP Urband Microcell model \cite{3GPP_2017_TS}. The large scale fading coefficient is given as 
\begin{equation}
\beta_{l,k}[dB] = -30.5 -36.7 {\sf log}_{10}\left( \frac{d_{l,k}}{1 {\rm m}}  \right) +F_{l,k},
\end{equation}
where $d_{l,k}$ is the distance between AP $l$ and UE $k$ and $F_{l,k} \sim \mathcal{N}_\mathbb{C} (0,4^2)$ is the shadow fading. The shadowing terms from an AP to different UEs are correlated as $\Ex [F_{l,k}F_{l,i}] = 4^2 2^{- \delta_{k,i}/9 {\rm m}}$, where $\delta_{k,i}$ is the distance between UE $k$ and UE $i$. The pilots are randomly indexed and assigned to all UEs. We employ $4$-QAM modulation scheme and set the signal-to-noise ratio (SNR) $25$ dB. We evaluate the performance depending on the  pilot-to-user ratio defined as $\tau/K$.  Finally, we compare the performance of our proposed detector with the CF-MRC \cite{2017HQNgo_TWC_CFvsSmall},  CF-MMSE \cite{2016ENayebi_Asilomar_CFMMSE}, and  CF-MMSE-SIC \cite{2020EBjornson_TWC_MakingCFcompetitive} detectors. 

\subsection{Convergence and Complexity Analyses}\label{ConvAndComplex}

We first analyse the convergence behaviour of the proposed CF-EP detector by plotting the average number of iterations needed to satisfy the MM condition in \eqref{eqEP:MM}. It can be seen from Fig. \ref{F2} that the CF-EP detector needs less than $6$ iterations to converge for various pilot-to-user ratios, $\tau / K$. 


As tabulated in Table 1, the computational complexity of the CF-EP detector is slightly higher than the MMSE detector depending on its number of iterations. The CF-EP requires on average $2-6$ iterations to converge. Therefore, the complexity order of the CF-EP detector remains the same as the CF-MMSE detector.  Note that the detector will be implemented at the CPU, which is assumed to be equipped with a high performance computer. Therefore, a minor increase in the computational complexity will not be a critical issue.

\begin{table}\small
  \begin{tabular}{| c| c|}
  \hline
 Detector 			&     	Complexity		\\ 
     	 \hline
  \hline
    		CF-MRC 			& 		$\mathcal{O} (NK)$ 				\\
     	 \hline
    		CF-MMSE		& 		$\mathcal{O} (N^2K)$ 	 			\\
    		\hline
     	CF-MMSE-SIC & 		$\mathcal{O} (\sum_{k=0}^{K-1}(N-k)^2(K-k))$   			\\
     	 \hline
     	CF-EP  		& 		$\mathcal{O} (N^2KT)$ 	 		\\
     	 \hline
  \end{tabular}  \label{T1}
  \caption{Computational complexity comparison}
\end{table}

\subsection{Performance Analysis}

\begin{figure}
\centering
\subfloat[The BER performance]
{\includegraphics[scale=0.27]{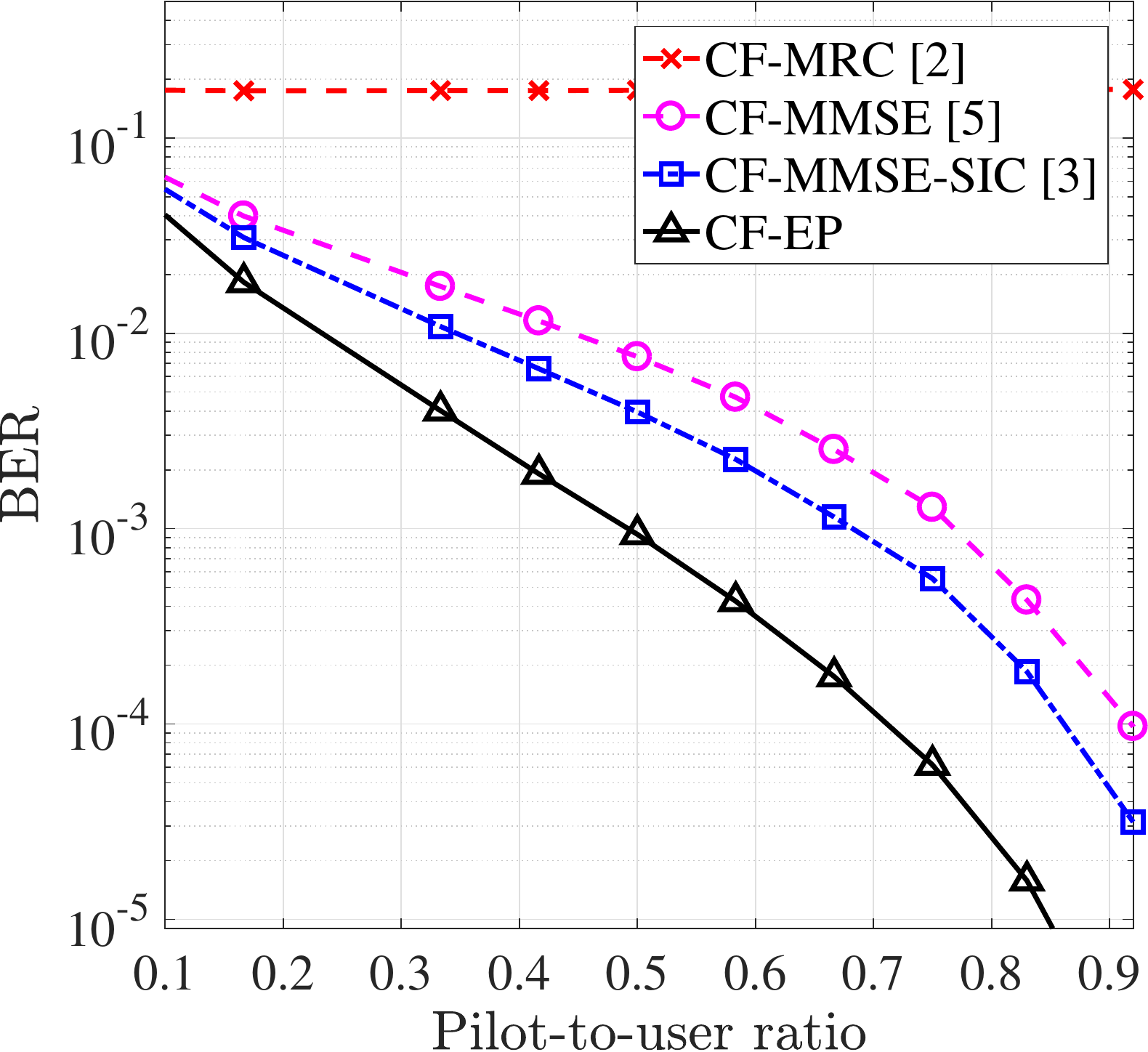}}\hfill \hfill
\centering
\subfloat[The sum SE performance]
{\includegraphics[scale=0.27]{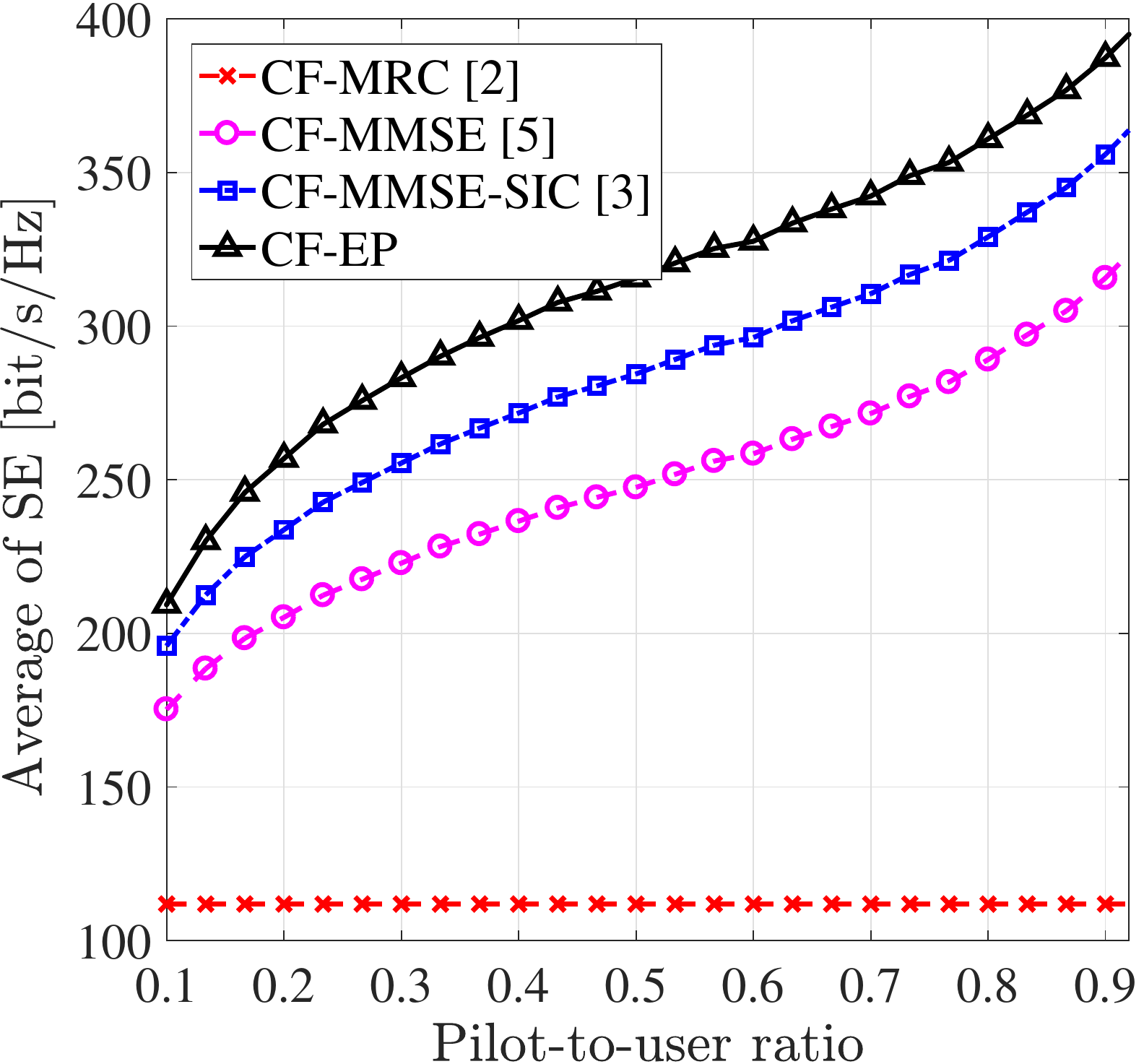}}
\caption{The BER and sum SE performances of CF-MRC \cite{2017HQNgo_TWC_CFvsSmall}, CF-MMSE \cite{2016ENayebi_Asilomar_CFMMSE}, CF-MMSE-SIC \cite{2020EBjornson_TWC_MakingCFcompetitive}, and CF-EP detectors with $L=100, K=60$, $M=4$, SNR = $25$ dB}
\label{F4}
\end{figure}

To analyse the proposed detector's performance, we plot the BER and sum SE over the random users  locations versus the pilot-to-user ratio.
Fig. \ref{F4}a demonstrates that the CF-EP detector outperforms the CF-MMSE-SIC  and the CF-MMSE detectors in terms of BER, for all pilot-to-user-ratio $\tau/K$. More specifically, the CF-EP detector can achieve BER $10^{-4}$ with around $ 25 \% $ lower pilot-to-user ratio than the CF-MMSE and CF-MMSE-SIC detectors.  
In Fig. \ref{F4}b, the CF-EP detector demonstrates $26 \% $ and  $ 11 \% $ sum SE improvements compared to the CF-MMSE and CF-MMSE-SIC detectors, respectively. 
 From these facts, we conclude that the developed EP based detector, dealing with the pilot contamination  impact, achieves a significant performance gain compared to the existing CF M-MIMO detectors.


\section{Conclusion}

We propose a novel EP based detector for CF M-MIMO systems, namely the CF-EP detector.
 Our simulation results show that the BER and sum SE performances of the CF-EP detector are better than those of the state-of-the-art CF M-MIMO detectors with comparable computational complexity.

\section*{Acknowledgment}
This research was supported by the research training program stipend from The University of Sydney. The work of Branka Vucetic was supported in part by the Australian Research Council Laureate Fellowship grant number FL160100032.
%

{\renewcommand{\baselinestretch}{1.1}
\begin{footnotesize}
\bibliographystyle{IEEEtran}
\bibliography{IEEEabrv,myBib}
\end{footnotesize}}

\end{document}